# *Temporal chirp, temporal lensing and temporal routing via space-time interfaces*

*Victor Pacheco-Peña[1], Mathias Fink[2] and Nader Engheta[3]*

[1] *School of Mathematics, Statistics and Physics, Newcastle University, Newcastle Upon Tyne, NE1 7RU, United Kingdom*
[2] *Institut Langevin, ESPCI Paris, CNRS, PSL University, 1 rue Jussieu, 75005 Paris, France*
[3] *Department of Electrical and Systems Engineering, University of Pennsylvania, Philadelphia, PA 19104, USA*

email: victor.pacheco-pena@newcastle.ac.uk, mathias.fink@espci.fr, engheta@seas.upenn.edu

A time interface (a rapid change of the constitutive parameters of a material in time), applied within an unbounded medium where a wave travels, can enable frequency conversion, and is considered the temporal analogue of a spatial interface between two materials. Here, we study light-matter interactions in four dimensions, 4D (space, *x,y,z,* and time, *t*), by exploring the implications of applying time interfaces not to the entire space where a wave travels, but to certain regions of space in order to create spatial interfaces in time. Different configurations such as induced perpendicular, parallel, and oblique spatial interfaces via a temporal interface are discussed. It is shown how such four-dimensional combination of temporal and spatial interfaces can enable interesting features such as the 4D *generalized Snell's law* and *temporal chirp, temporal lensing,* and *temporal routing* of electromagnetic waves. Such exotic possibilities may provide new ways to manipulate light-matter interactions via a combination of temporal and spatial interfaces.

It is well known that wave-matter interactions can be tailored by engineering three-dimensional spatial ($x,y,z$) inhomogeneities of materials along the path of wave propagation [1]. In seeking alternative ways to manipulate waves, the scientific community has reported great progress in fields such as metamaterials, plasmonics, and photonics [2–4], enabling groundbreaking applications in diverse areas such as radars [5–7], imaging [8,9], and computing [10–13], to name a few. While all these applications have been developed using materials with time-invariant parameters such as relative permittivity, $\varepsilon$, and relative permeability, $\mu$, the focus has now been shifting towards materials with such parameters being time-dependent. In so doing, a higher degree of freedom is enabled by introducing *time* as an additional variable in material parameters, opening new paradigms for a full control of electromagnetic (EM) wave propagation in space and time [14–17].

An interesting type of time-modulated media is time interfaces where $\varepsilon$ and/or $\mu$ of the *whole* medium are rapidly changed in time (with timescales smaller than the period $T$ of the propagating wave) while the wave is in it [18,19]. These time interfaces have been recently proposed for interesting applications such as filters [20,21], temporal inverse and space-time Fresnel prisms [22,23], temporal aiming and Brewster angle [24,25], meta-atoms in space and time [26,27], temporal and spatio-temporal gratings [28,29], with groundbreaking experimental demonstrations being reported for water waves [30], microwaves [31] and optical frequencies [32,33]. When applying a time interface at a time $t = t_0$, provided that the impedance is changed from $(Z_1, t < t_0)$ to $(Z_2, t > t_0)$ with $Z_1 \neq Z_2$, time-refracted *forward* (FW) and time-reflected *backward* (BW) waves are generated [18,34]. While time interfaces are applied within an unbounded medium, a combination of spatial and temporal interfaces (space-time interfaces) have recently been explored in the realm of surface waves on time-dependent metasurfaces [35] and effective medium theory [36–38] (where both spatial and temporal interfaces have dimensions smaller than the incident wavelength, $\lambda$, and $T$ of the propagating wave, respectively).

Inspired by the exciting possibilities that time-dependent media can offer, here we explore wave propagation in a medium where spatial interfaces are created in time via an induced time interface at a time $t = t_0$. We first discuss the physical phenomena that arise when inducing in time a single perpendicular, parallel, or oblique spatial interface within an unbounded medium (Fig. 1), providing physical insights in terms of field distributions and space-time mapping of wave propagation. Then, more complex configurations are presented, showing how the combination of multiple space-time interfaces can lead to phenomena such as the temporal equivalent of generalization of the Snell's law in 4D (*x,y,z,t*) and temporal chirp, temporal lensing, and temporal routing demonstrating the potentials of space-time interfaces.

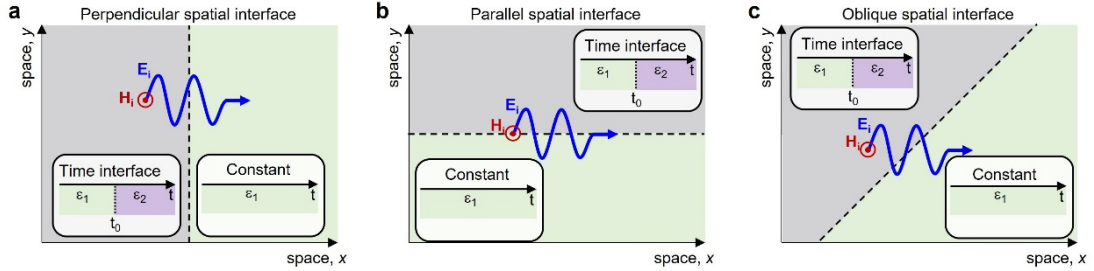

Figure 1. A wave propagates within an unbounded medium for times $t < t_0$. At $t = t_0$, a temporal interface is induced within a half-space of an unbounded medium by changing $\varepsilon$ from $\varepsilon_1$ to $\varepsilon_2$. This enables the creation of a (a) perpendicular, (b) parallel or (c) oblique spatial interface. The spatial region filled with a time dependent $\varepsilon(t)$ is depicted in gray while the green spatial region represents the half-space where $\varepsilon$ is constant in time. The schematic representation of $\varepsilon$ in the time domain for each spatial region is shown as insets in each panel.

Consider the scenarios shown in Fig. 1. For times $t < t_0$, a monochromatic EM wave is traveling within an unbounded (homogeneous and isotropic) medium having a relative permittivity value of $\varepsilon_1$. The incident wave is polarized along *y* ($E_y$) and propagates along *x*. At $t = t_0$, the $\varepsilon$ of just one half-space is changed to $\varepsilon_2$ (all values real and larger than 1, with $\mu_1 = \mu_2 = 1$) while the other half-space remains with the same initial value $\varepsilon_1$ (constant $\varepsilon$). The boundary of the spatial region filled with a time-dependent $\varepsilon(t)$ (gray) can be chosen to be perpendicular, parallel, or oblique with respect to the direction of propagation of the wave such that different spatial interfaces can be created (see Fig. 1). To begin with, let us study the case where a perpendicular spatial interface is created using a time interface. This configuration is shown in Fig. 1a and Fig. 2a for completeness. To evaluate it, we make use of the same configurations used in [38] but now considering that not the *whole* but just half-space is time-modulated via $\varepsilon(t)$. Moreover, the incident signal is switched off when inducing a time interface in order to better appreciate the effects of the temporal and spatial interfaces on the wave that is already in the region.

The numerical results of the electric field distribution ($E_y$) as a function of space and time are shown in Fig. 2b where an EM wave is traveling within an unbounded medium for $t < t_0$ with $\varepsilon_1 = 1$ (see supplementary materials for an animation of this case). At $t \sim 28.35T$ (*T* being the period of the monochromatic wave for $t < t_0$) a temporal interface is applied to the left half-space in Fig. 2a (Medium A shown in Fig. 2b) where $\varepsilon$ is changed to $\varepsilon_2 = 10$, while the right half-space (Medium B) remains unchanged. To better analyze these results, we provide in Fig. 2c a space-time map of all the temporal and spatial waves produced by the temporal and induced spatial interfaces. After the time interface is applied, there is a *remnant* signal that continues its propagation in Medium B with the same frequency as for times $t < t_0$, i.e., $f_1$, since $\varepsilon$ is constant in this region (no change of frequency or wavelength $\lambda_0$). For Medium A, however, the generated FW and BW waves have a new frequency $f_2 = f_1\sqrt{\varepsilon_1/\varepsilon_2}$ ($\lambda_0$ and wavenumber $\boldsymbol{k}$ do not change). The BW wave simply travels along the negative $x$ axis, i.e., $x/\lambda_0 < 0$ within Medium A with a lower speed (reduced $T$) since $\varepsilon_2 > \varepsilon_1$. The FW wave travels towards the positive $x$ direction, and when it reaches the induced spatial boundary at $x/\lambda_0 = 0$, spatial transmission ($T_x$) and reflection ($R_x$) occurs. From Fig. 2b, this can be seen as a change in wavelength in Medium B compared to the wavelength in Medium A which remains unchanged for all times.

A natural question arises: Could we exploit such time-induced perpendicular spatial interfaces for more complex wave-matter interactions? Imagine the spatial scenario shown in Fig. 2d (note that for this example the axes have been changed compared to Fig. 2a to better compare it with its space-time equivalent, as it will be shown below). As is known, if an incident oblique incident wave with wavenumber $\boldsymbol{k_1}$ ($k_{x1}, k_{z1}$) reaches a spatial boundary between two materials, a reflected (not shown) and a refracted wave will travel within the first and second medium, respectively. The direction of the latter wave is defined by Snell's law which requires $k_{x2}$ in medium 2 to be the same as $k_{1x}$ in medium 1, i.e., $k_{x2}(x) = k_{x1}$, leading to the fact that the entire refracted wave propagates along a single direction [1]. Interestingly, as shown in [39] one can generalize Snell's law in space by placing properly engineered subwavelength scatterers of different geometries/dimensions along the $x$ axis on the spatial interface. These scatterers cause an extra phase shift $\varphi(x)$ between the two sides of the interface. In so doing, a variation of phase $\varphi(x)$ can be introduced such that the wavenumber can be locally changed to $k_{x2}(x) = k_{x1} + \partial\varphi(x)/\partial x$. This generalized Snell's law in space has opened new avenues in the field of metasurfaces and metamaterials to arbitrarily manipulate fields and waves [39]. Here we propose a way to further generalize such concept to four dimensions – space and time ($x,y,z,t$) – by creating spatial interfaces via time interfaces. Our approach is schematically shown in Fig. 2e where multiple perpendicular spatial interfaces are induced at the same time using time interfaces with different values of $\varepsilon$, all applied at $t = t_0$. In so doing, as shown in Fig. 2e, an array of spatial layers is created in time with the frequency of the signal inside them being different, as expected ($f_{2\_lm} = f_1\sqrt{\varepsilon_1/\varepsilon_{lm}}$, $m = 1,2,3 ...$ where $m$ being the spatial layer number). This means that for $t < t_0$ the angular frequency of the wave in the entire space is $\omega_1$ and for $t > t_0$ the converted frequency just due to the temporal interfaces is a function of position $\omega_2(x)$, which can be considered as the spatio-temporal analogue of the generalized Snell's law in space, $k_{x2}(x)$. Examples of this concept are shown in Fig. 2f,g where 10 perpendicular spatial layers of thickness $2\lambda_1$ are induced at $t = t_0$ (see caption for further details). These two examples consider the scenarios where i) $\varepsilon$ is time-dependent, $\varepsilon(t)$, while µ is constant ($µ_1 = µ_{lm} = 10$) and ii) µ is time-dependent, µ(t), while $\varepsilon$ is constant ($\varepsilon_1 = \varepsilon_{lm} = 10$). The values of $\varepsilon(t)$ for case i) are shown on the left panels of Fig. 2f,g [the values of µ(t) are the same so they are not repeated in the figure]. The numerically simulated $E_y$ as a function of time, recorded at a point in the spatial region where $\varepsilon$ and µ do not change (a point in the right green region in the middle panel of Fig. 2e), is shown in Fig. 2f,g for both, cases i) and ii), respectively. As observed, the resulting wave traveling towards the region of constant $\varepsilon$ and µ has oscillations with different amplitudes and time durations, with the latter feature effectively being a *temporal chirped* signal. Such *temporal chirp* can be tailored by modifying the values of $\varepsilon_{lm}$ or $µ_{lm}$ of each induced spatial layer, see Fig. 2f,g. This can be explained via the FW and BW waves created in each induced spatial layer. The amplitudes of the FW and BW waves as well as the converted frequency, $f_{2\_lm}$, are larger in spatial layers where $\varepsilon(t)$ or µ(t) are changed to smaller values ($\varepsilon_{lm}, µ_{lm} < \varepsilon_1, µ_1$) [18]. This can be seen in the recorded final *temporal chirped* signals from the two examples in Fig. 2f.g. Larger amplitudes and shorter time oscillations (larger $f_{2\_lm}$) appear at the extreme ends (left and right) or at the center of the time window from Fig. 2f,g, respectively (generated by layers with $m = 1, 10$ and $m = 5, 6$ respectively), demonstrating the potential to achieve a tailored *temporal chirped* signal. See also a supplementary video for the case shown in Fig. 2g.

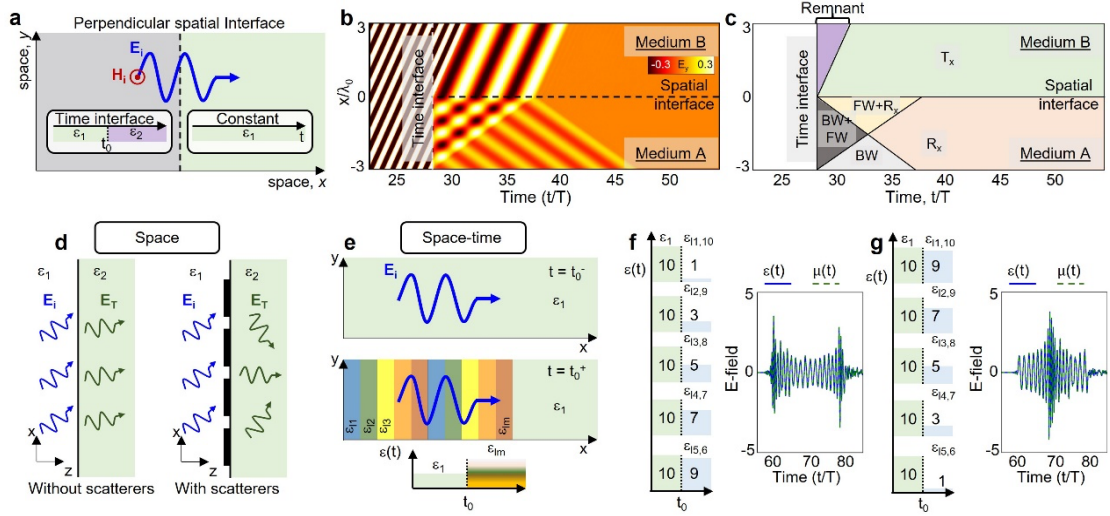

Figure 2. (a) Schematic of an induced perpendicular spatial interface created by a temporal interface. (b,c) space-time diagrams for $E_y$ (calculated at $y = 0$) and space-time waves generated when an initial wave is traveling in an unbounded medium with $\varepsilon_1 = 1$, and a temporal interface is applied at $t = 28.35T$, $\varepsilon$ is changed to $\varepsilon_2 = 10$. The μ remains unchanged in time with $\mu_1 = \mu_2 = 1$. (d) Generalized Snell's law in space (three dimensions, $x,y,z$). (e) 4D generalized Snell's law (space-time, $x,y,z,t$) where multiple perpendicular spatial layers are created by applying different temporal interfaces simultaneously. (f,g) Numerical simulation results of the recorded $E_y$ for the FW wave (recorded in the spatial region where $\varepsilon$ and μ are time-independent, right green region from Fig. 2e, middle) after applying different temporal interfaces of different values simultaneously along the spatial $x$ axis. A total of 10 layers are produced with a thickness of $2\lambda_1$ ($\lambda_1$ is the wavelength inside the whole medium having $\varepsilon$ and μ values as those from $t < t_0$). The values of $\varepsilon(t)$ for each induced spatial layer are shown on the left panels considering $\mu_1 = \mu_{lm} = 10$. The opposite case is also evaluated where μ is time-dependent, $\mu(t)$, and has the same values as those shown for $\varepsilon(t)$ in panels f,g while $\varepsilon_1 = \varepsilon_{lm} = 10$ for all the spatial layers. The results when changing $\varepsilon$ or changing μ are shown as solid blue or dashed green lines, respectively.

Let us now study the case where a parallel spatial interface is induced via a temporal interface. This scenario is shown in Fig. 1b and Fig. 3a. As in Fig. 2, half-space of an unbounded medium is filled with a time-dependent $\varepsilon(t)$ (Medium A, top) while the other half is kept constant ($\varepsilon$ does not change, Medium B, bottom). The numerical results (using the same constitutive parameters as those from Fig. 2b for $t < t_0$) of the $H_z$ distribution as a function of space and time along with a schematic of the generated space-time waves are shown in Fig. 3b,c, respectively (in this example, $\varepsilon$ for Medium A is changed to $\varepsilon_2 = 2$ at $t = t_0$). As observed, for $t > t_0$, a FW and a BW wave are created within Medium A traveling with a different frequency $f_2$ (as defined above), while a remnant wave continues its propagation within Medium B with the same initial frequency $f_1$, as expected. Moreover, a radiated signal is created along the full spatial interface ($y = 0$). Intriguingly, this radiated field is more evident in Medium A ($y/\lambda_0 > 0$) where some ripples can be seen for times between $30T < t < 35T$. Remarkably, this is in fact the temporal version of the interfacial antenna scenario explored in [40] where the radiation of a line source placed at the spatial interface between two dielectrics was studied. In that work, it was shown that the radiated field is stronger within the denser material (larger $\varepsilon$) and this radiation has its peak along a specific angle defined by the values of $\varepsilon$ from both media. This, indeed, is analogous to what is happening in the space-time scenario shown in Fig. 3b,c, where the radiated field produced along the induced spatial interface is stronger in Medium A as $\varepsilon_2 = 2$ in this medium, while it is $\varepsilon_1 = 1$ in medium B. See supplementary materials for an animation of this scenario.

As an application of such induced parallel spatial interface, we propose the concept of a *temporal lens*. Consider the case shown in Fig. 3d. As is well known, the shape and materials of a conventional (spatial) lens should be designed to manipulate the phase of the output signal such that an incoming collimated beam upon passing through the lens converges to the focal point. Similarly, for our proposed *temporal lens*, as shown in Fig. 3e, one can induce in time an array of different but simultaneously created parallel spatial layers in which the frequencies and velocities of the waves are different. Importantly, for our *temporal lens*, the $\varepsilon$ and/or $\mu$ of each induced parallel layer created at $t = t_0$ should be chosen such that the new frequency and phase velocity in the central layers should be smaller than those of the side-layers. i.e., $\varepsilon$ and/or $\mu$ for the central parallel spatial layers should be larger than those in the side layers for $t > t_0$. In so doing, as described in Fig. 3b, each spatial interface between parallel layers will create radiated fields which will travel towards the denser parallel layers until all converge at $y = 0$, along the entire spatial $x$ axis. An example is shown in Fig. 4g (and an animation is also available as a supplementary material) where $\varepsilon$ is modified at $t = 20T$ from $\varepsilon_1 = 10$ (for the whole medium where the wave is initially traveling) to different values along the $y$ axis to create 10 induced parallel spatial layers each of height $\lambda_1$. In this process, $\mu$ is kept unchanged ($\mu_1 = \mu_{lm} = 1$). Faster/slower FW waves are created within each induced parallel spatial layer depending on the values of $\varepsilon_{lm}$ used, as expected, with slower FW waves for the central layers. As is shown, a "line" focus is produced along the whole spatial $x$ axis at $y = 0$. This signal will then diverge in time (as shown in the supplementary video of this case), similar to the divergence of waves after focusing in space in a conventional convex lens (Fig. 3d), demonstrating the performance of the proposed *temporal lens* concept.

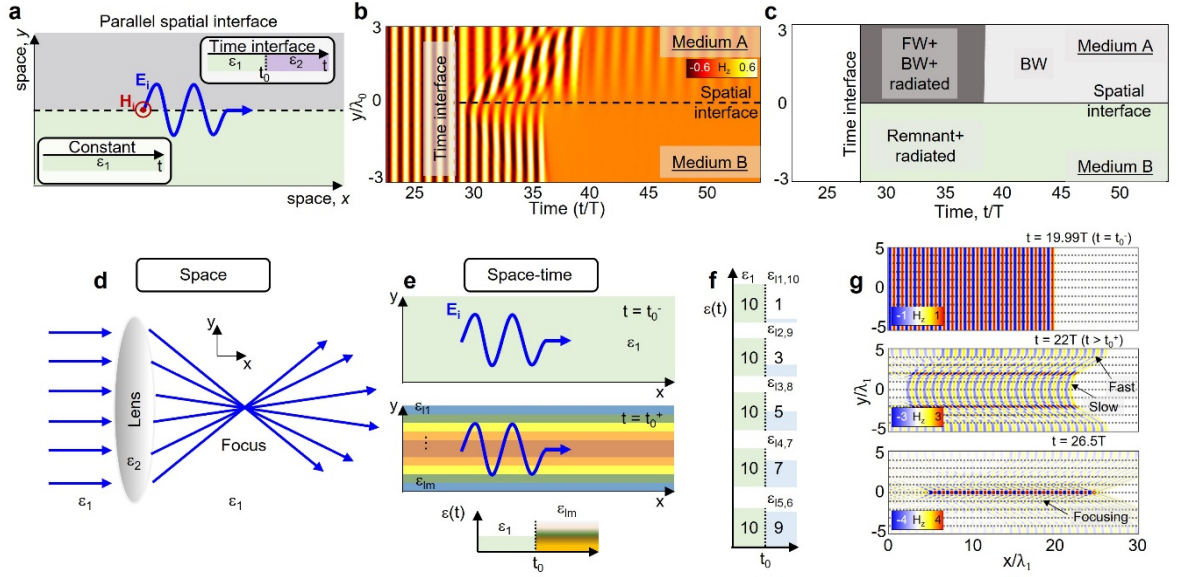

FIG. 3 (a) Schematic of an induced parallel spatial interface created at a temporal interface. (b,c) space-time diagrams showing the $H_z$ field distribution (at $x = 0$) and space-time waves. The values of $\varepsilon_1$, $t_0$ and μ are the same as Fig. 2b. $\varepsilon$ is changed from $\varepsilon_1 = 1$ to $\varepsilon_2 = 2$ at $t = t_0$. (d) Schematic of a conventional (spatial) lens. (e) *Temporal lens* concept in four dimensions. Multiple parallel spatial layers are created by applying different temporal interfaces simultaneously. (f,g) Numerical simulation results of the $H_z$ field distribution at different times showing how a "line" focus is produced after inducing the spatial interfaces via temporal interfaces (bottom panel from g). A total of 10 layers are used with a thickness of $\lambda_1$ (as the wavelength inside the whole medium having $\varepsilon$ and μ values as those from $t < t_0$). The values of $\varepsilon(t)$ for each layer are shown on (f) with $\mu_1 = \mu_{lm} = 1$.

As a final scenario, consider the spatial interface shown in Fig. 1c and Fig. 4a. As in the previous examples, a time interface is induced in one half-space of the unbounded medium, where a wave is traveling, to create an oblique spatial interface. For an example of an application of such space-time configuration, we propose the *temporal router* shown in Fig. 4b-d where an EM wave can be redirected. Here, a wave is propagating within the whole medium (a Gaussian signal in this case) with a frequency $f_1$ and $\varepsilon_1 = \mu_1 = 1$ (Fig. 4b). At $t = t_0 = 59.7T$, a temporal interface is induced to the left half-space (Medium A) to create an oblique spatial boundary of angle $\theta = 45°$. To do this, $\varepsilon$ of Medium A is changed to $\varepsilon_2 = 10$ while $\varepsilon$ for Medium B (right half-space) remains unchanged, $\varepsilon_1$. As shown in Fig. 4c, for a time $t > t_0^+$, a remnant wave travels within Medium B with the same frequency and wavelength as the initial frequency and wavelength, as expected and discussed in Fig. 2,3. In medium A, a FW and BW waves with the converted frequency are created. When the FW wave reaches the oblique spatial boundary, nothing is transmitted into Medium B, but it is perfectly reflected with propagation along the *y* axis, completely changing its direction of propagation due to the incident angle ($\theta = 45°$) being larger than the critical angle $\theta = \mathrm{asin}\left(\sqrt{\varepsilon_B/\varepsilon_A}\right) = \mathrm{asin}\left(\sqrt{\varepsilon_1/\varepsilon_2}\right)$ (with $\varepsilon_B$ and $\varepsilon_A$ as the relative permittivity values for Medium B and A, respectively, after inducing the time interface in Medium A). This is due to the total internal reflection that occurs after temporal interface is applied. See also Fig. 4d. A supplementary video is also available to further observe this performance.

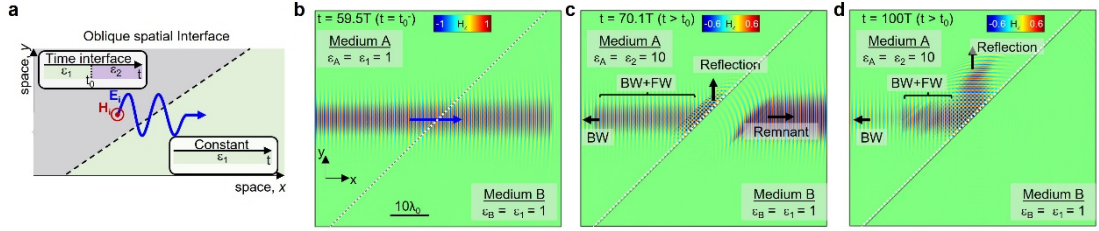

FIG. 4 (a) Inducing an oblique spatial interface using a temporal interface. (b-d) $H_z$ field distribution at times before (b) and after (c,d) applying a temporal interface to the left medium of (a). µ is not time-dependent.

In conclusion, we have discussed how temporal interfaces can be exploited to create different types of spatial interfaces (perpendicular, parallel, oblique), providing various wave-matter interactions for the spatial and temporal signals present in such scenarios. Different wave characteristics involving such temporally induced spatial interfaces have been discussed including the 4D generalization of Snell's law for temporal chirp, temporal lensing as well as temporal routing of EM waves. These results may offer new opportunities and directions for controlling EM waves in multiple dimensions via the implementation of temporal interfaces and the combinations of multiple types of induced spatial interfaces (perpendicular, parallel and oblique).

## Acknowledgements

V. P.-P. would like to thank the support of the Leverhulme Trust under the Leverhulme Trust Research Project Grant scheme (RPG-2020-316). M. F. and N. E. acknowledge partial support from the Simons Foundation/Collaboration on Symmetry-Driven Extreme Wave Phenomena (grant 733684 (N.E.) and grant 733686 (M.F.). For the purpose of Open Access, V. P.-P. has applied a CC BY public copyright license to any Author Accepted Manuscript (AAM) version arising from this submission.

## Conflicts of interests

The authors declare no conflicts of interests.

## Data availability

The datasets generated and analyzed during the current study are available from the corresponding author upon reasonable request.